\documentclass[twocolumn,english,preprintnumbers,amsmath,amssymb]{revtex4}

\usepackage{graphics,graphicx}
\usepackage{subfigure}
\usepackage{dcolumn}
\usepackage{bm}

\begin{document}

\title{X-ray scattering study of the spin-Peierls transition and soft phonon behavior in TiOCl}

\author{E.T. Abel$^1$}
\author{K. Matan$^1$}
\author{F.C. Chou$^{2}$}
\author{E.D. Isaacs$^3$}
\author{D.E. Moncton$^1$}
\author{H. Sinn$^4$}
\author{A. Alatas$^4$}
\author{Y.S. Lee$^{1*}$}

\affiliation{$^1$Department of Physics, Massachusetts Institute of
Technology, Cambridge, MA 02139}

\affiliation{$^2$Center for Condensed Matter Sciences, National
Taiwan University, Taipei 10617, Taiwan}

\affiliation{$^3$Center for Nanoscale Materials, Argonne National
Laboratory, Argonne, IL 60439}

\affiliation{$^4$Advanced Photon Source, Argonne National
Laboratory, Argonne, IL 60439}

\

\date{\today}

\begin{abstract}
We have studied the $S=1/2$ quasi-one-dimensional antiferromagnet
TiOCl using single crystal x-ray diffraction and inelastic x-ray
scattering techniques.  The Ti ions form staggered spin chains which
dimerize below $T_{c1} = 66$~K and have an incommensurate lattice
distortion between $T_{c1}$ and $T_{c2} = 92$~K.  Based on our
measurements of the intensities, wave vectors, and harmonics of the
incommensurate superlattice peaks, we construct a model for the
incommensurate modulation.  The results are in good agreement with a
soliton lattice model, though some quantitative discrepancies exist
near $T_{c2}$.  The behavior of the phonons has been studied using
inelastic x-ray scattering with $\sim 2$ meV energy resolution. For
the first time, a zone boundary phonon which softens at the
spin-Peierls temperature $T_{SP}$ has been observed.  Our results
show reasonably good quantitative agreement with the Cross-Fisher
theory for the phonon dynamics at wave vectors near the zone
boundary and temperatures near $T_{SP}$. However, not all aspects of
the data can be described, such as the strong overdamping of the
soft mode above $T_{SP}$. Overall, our results show that TiOCl is a
good realization of a spin-Peierls system, where the phonon
softening allows us to identify the transition temperature as
$T_{SP}=T_{c2}=92$~K.
\end{abstract}
\maketitle

\section{\label{sec:level1} Introduction}

Understanding the properties of low-dimensional spin-1/2 systems
remains a central challenge in condensed matter physics.  In
quasi-one-dimensional materials, a variety of ground-states can be
stabilized, depending on the subtle interactions beyond the dominant
magnetic exchange along the spin chain.  For example, the low-energy
physics of the $S=1/2$ Heisenberg spin chain\cite{Bethe31,Hulthen38}
becomes modified once coupling to the three-dimensional crystal
lattice is taken into account.  In the presence of spin-phonon
coupling, the ground-state is dimerized, corresponding to the
spin-Peierls state.\cite{Pytte74,CrossFisher79}  If inter-chain
magnetic interactions are significant, they can promote N$\acute{\rm
e}$el order as an alternate ground state. Even though many
quasi-one-dimensional $S=1/2$ spin chain materials have been studied
so far, only a relatively small fraction are observed to have
spin-Peierls ground states. The effects of the spin-phonon coupling
in these and other low dimensional quantum spin systems are
important topics for further investigation.

The spin-Peierls transition has been observed in many organic charge
transfer compounds, such as
TTF-CuBDT.\cite{Jacobs76,Moncton77,Huizinga,Bray75,Visser83,Pouget01}
In these systems, a static lattice dimerization has been measured in
diffraction experiments.  In addition, x-ray measurements of diffuse
scattering suggested the presence of a soft phonon associated with
the transition\cite{Moncton77}. However, direct measurements of the
relevant lattice dynamics are lacking due to the unavailability of
large crystals for inelastic neutron scattering. The observation of
a spin-Peierls transition in the inorganic compound
CuGeO$_3$\cite{Hase93} rekindled interest in spin-Peierls research,
as large single crystals could be readily grown. However, it is not
clear whether CuGeO$_3$ is a good realization of a conventional
spin-Peierls system. The susceptibility does not follow the expected
behavior of a spin chain with nearest neighbor
coupling\cite{Hase93}, and it is believed that significant next
nearest neighbor coupling along the chain must be taken into
account.\cite{Riera95}  In fact, the ratio of the nearest neighbor
to next nearest neighbor exchange interaction is close to the
critical value for spontaneous formation of a spin gap independent
of a lattice distortion.\cite{Riera95} Neutron scattering
measurements also reveal a significant magnetic coupling between
neighboring chains\cite{Regnault96}.

Another puzzle in CuGeO$_3$ is the lack of an observed phonon
softening at the transition to the dimerized state. Cross and Fisher
developed a theory which includes the spin-phonon coupling and
predicts a softening of a spin-Peierls active phonon to zero
frequency at $T_{SP}$.\cite{CrossFisher79}  Neutron studies were
unable to detect the presence of this soft phonon, and, in fact, and
a hardening of phonon modes was observed\cite{Lorenzo94,Braden02}.
Recent theoretical analysis suggests that CuGeO$_3$ may fall within
the ``anti-adiabatic'' regime (characterized by a large phonon
frequency relative to the magnetic interaction energy) where a
phonon hardening would be
expected\cite{Gros98,Holicki,Orignac04,Citro}.  The lattice dynamics
associated with the more conventional ``adiabatic'' regime have yet
to be measured in detail.

\begin{figure}[t!]
  \centering
  \includegraphics[width=0.35\textwidth]{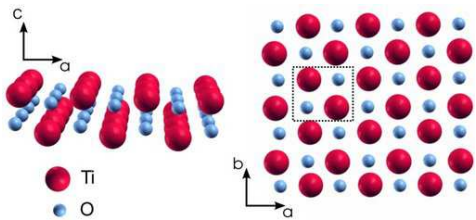}\\
  \caption{\label{fig:tioclstructure}
Structure of TiOCl at high temperatures in the undistorted state.
Left: Schematic of a Ti-O layer showing the Ti ion chains along the
$b$-direction as well as the positions of the oxygens. Right:
Schematic of the $ab$-plane which more clearly shows the staggered
arrangement of adjacent Ti chains. The dashed lines highlight the
unit cell.}
\end{figure}

Recently, a new $S=1/2$ spin chain material has been discovered: the
transition metal oxide TiOCl.  The structure of TiOCl is
orthorhombic (Pmmn), consisting of TiO planes (shown in
Fig.~\ref{fig:tioclstructure}) separated by Cl layers. The Ti$^{3+}$
ions are in the $d^1$ electronic configuration with $S=1/2$.
Initially, TiOCl was thought to be an attractive candidate for the
``resonating valence bond'' state\cite{wilson93}, but it has since
been found to be primarily a one-dimensional magnetic system. Band
structure calculations indicate that TiOCl is a Mott insulator with
the lowest occupied Ti $d_{xy}$ orbitals pointed toward each other
along the crystallographic $b$-direction.\cite{Seidel03,Hoinkis} The
high temperature magnetic susceptibility is well described by the
Bonner-Fisher curve, indicating a nearest neighbor magnetic exchange
of $J \simeq 660$~K. \cite{Seidel03}  In the TiOCl crystal
structure, the spin chains are staggered--every other Ti chain is
displaced by half a lattice spacing along the $b$-direction.  Hence,
the inter-chain magnetic interaction is frustrated.

Upon cooling, the susceptibility falls away from the Bonner-Fisher
curve around $T\simeq150$~K and exhibits anomalies at $T_{c2}=92$ K
and $T_{c1}=66$ K.\cite{Seidel03}  The drop at 150~K has been
attributed to the opening of a pseudo-gap \cite{Imai03,Clancy}, the
anomaly at 92 K corresponds to the onset of an incommensurately
modulated state
\cite{Imai03,Abel04,Rueckamp05,Krimmel06,Schoenleber06}, and the one
at 66 K corresponds to a commensurate modulation\cite{Shaz05}.
Similar behavior has been observed in the isostructural compound
TiOBr.\cite{Lemmens_tiobr,vanSmaalen,Sasaki1,Sasaki2}  At first
sight, the observed transitions (such as the incommensurate phase)
suggest unconventional spin-Peierls behavior. In order to examine
the applicability of a spin-Peierls framework, several issues need
to be clarified, such as the nature of the incommensurate phase and
the lattice dynamics associated with the structural dimerization.

In this paper, we present x-ray diffraction and inelastic x-ray
scattering data on single crystal samples of TiOCl. As we shall
show, the material undergoes a spin-Peierls transition in which the
structural dimerization occurs in the presence of a soft-phonon at
the zone boundary.  The low temperature structural modulations are
discussed in some detail.  The format of the paper is as follows:
Section II outlines the crystal growth and thermodynamic
characterization of the samples.  Section III contains a discussion
of the low temperature structures observed with synchrotron x-ray
diffraction. Based on measurements of the superlattice peaks,
including higher harmonics, we develop a model for the
incommensurate structure described by a soliton lattice.  In Section
IV, the lattice dynamics are studied using high energy-resolution
inelastic x-ray scattering.  We observe a soft, zone boundary,
phonon mode which drives the spin-Peierls transition, and we compare
our data with the Cross-Fisher theory.  Finally, Section V contains
a discussion and summary of our results.

\section{Sample Growth and Characterization}
\begin{figure}[b]
  \centering
  \includegraphics[width=0.45\textwidth]{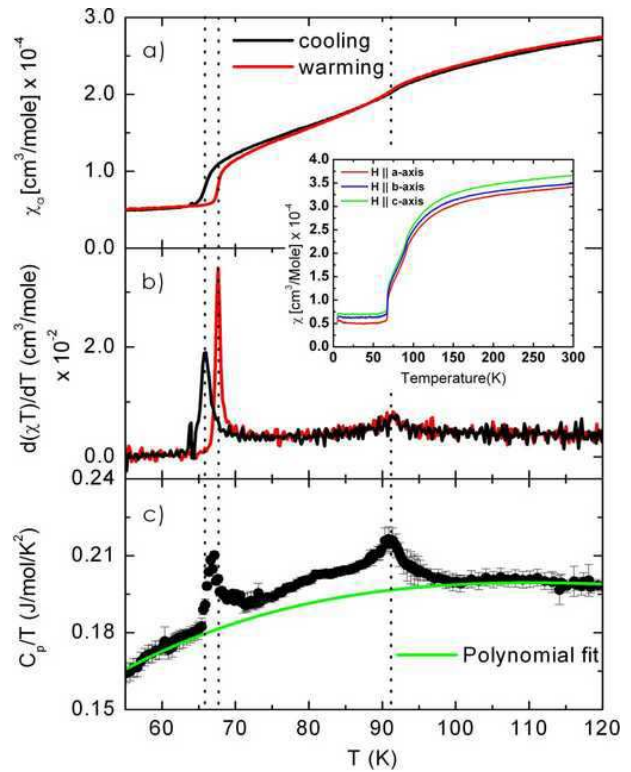}
  \caption{\label{fig:hc} Thermodynamic characterization of our sample.  (a)
    The magnetic susceptibility upon warming and cooling after the subtraction
    of a small Curie tail.  The sharp drops at $T_{c2}$=92 K and $T_{c1}$ = 66 K
    (cooling) indicate the two transition temperatures.  Inset: Susceptibility measured
    along the three crystallographic directions.  (b) The quantity $d(\chi T)/dT$
    (c) The specific heat $C_p(T)$.}
\end{figure}

The single crystals used in this work were grown by the vapor
transport method, using the procedure discussed
previously\cite{Seidel03}.  The typical size of an individual
crystal was about 50 mm x 50 mm x 20 $\mu m$ along the $a, b$ and
$c$ axes respectively.  The magnetic susceptibility was measured on
a sample composed of six single crystals co-aligned to within 4
degrees, for a total sample mass of about 8 mg.  The susceptibility
results (shown in Fig.~\ref{fig:hc}(a) and the inset) were performed
using a SQUID magnetometer in an applied magnetic field of 2 Tesla.
After the subtraction of a small isotropic Curie tail, the
susceptibility curves along the three crystallographic directions
are nearly identical (as shown in the inset), except for a small
temperature independent offset, most likely due to anisotropic
crystal field contributions.  This behavior further indicates the
Heisenberg nature of the spins.  Moreover, the isotropic drop in the
susceptibility is a signature for spin-singlet formation, as opposed
to N$\acute{\rm e}$el order.  Measurements of the susceptibility
(Fig.\ref{fig:hc}) upon warming and cooling show hysteretic behavior
around $T_{c1}=66$~K, which point to a first order phase transition.
Hysteresis is not observed around the transition at $T_{c2}=92$~K,
indicating that this transition is second order in nature. Below
T$_{c1}$, the susceptibility becomes temperature independent, as
seen previously\cite{Seidel03}, consistent with a fully formed
singlet state.  This apparent first order transition to the
spin-Peierls ground-state seems to differentiate TiOCl from the
other materials in which a second order spin-Peierls transition is
observed(\cite{Jacobs76,Moncton77,Huizinga,Bray75,Visser83,Pouget01,Hase93}).
However, as we will show in section VI, we find that it is actually
$T_{c2}=92$~K which defines the spin-Peierls transition.

Also shown in Fig.\ref{fig:hc}(b) is the quantity $d(\chi(T) T)/dT$
which should be proportional to the magnetic contribution to the
specific heat using the Fisher relation\cite{Johnston00,Fisher62}.
Indeed, the measured specific heat $C_{p}(T)$ in Fig.\ref{fig:hc}(c)
shows clear anomalies at $T_{c1}$ and $T_{c2}$ which find close
correspondence to anomalies in $d(\chi(T) T)/dT$. This suggests that
the origin of the anomalies in $C_{p}(T)$ have a significant
magnetic contribution.  A rough estimate of the magnetic
contribution can be obtained by integrating $C_{p}/T$ over $T$ in
order to calculate the change in entropy over the transition region.
At these relatively high temperatures, the phonon contribution
dominates the specific heat and must be subtracted.  We fit the
background in the vicinity of the transition region to a fourth
order polynomial denoted by the line in the Fig.~\ref{fig:hc}(c).
After background subtraction, the integration yields $\Delta
S=0.04(3)/NK_{B}$ for the entropy released in the vicinity of the
two transitions.  For comparison, one expects
$S/Nk_{B}\approx(2/3)(k_{B}T/J)$ for the entropy of a $S=1/2$
uniform chain at low-temperatures \cite{Johnston00}.  Using
$J/k_{B}=660$~K\cite{Seidel03} and taking $T$ to be an average of
$T_{c1}$ and $T_{c2}$ yields a rough upper bound for the change in
magnetic entropy of $\Delta S/Nk_{B} \simeq 0.08$.  In a previous
specific heat study, Hemberger {\em et al.}\cite{Hemberger}
estimated a value of $0.12$ for $S/Nk_{b}$, obtained by integrating
over a wider temperature range.  Our estimate is only meant to
account for the changes in the near vicinity of $T_{c1}$ and
$T_{c2}$, since the subtracted background must also contain a weakly
temperature dependent magnetic contribution.  Within the errors, our
results suggest a significant portion of the entropy change at the
transitions is magnetic in origin.

\section{\label{sec:xray}Single Crystal X-ray Diffraction}
\subsection{Superlattice Peaks}
\label{sec:commensurate}

\begin{figure}[t!]
  \centering
  \includegraphics[width=0.45\textwidth]{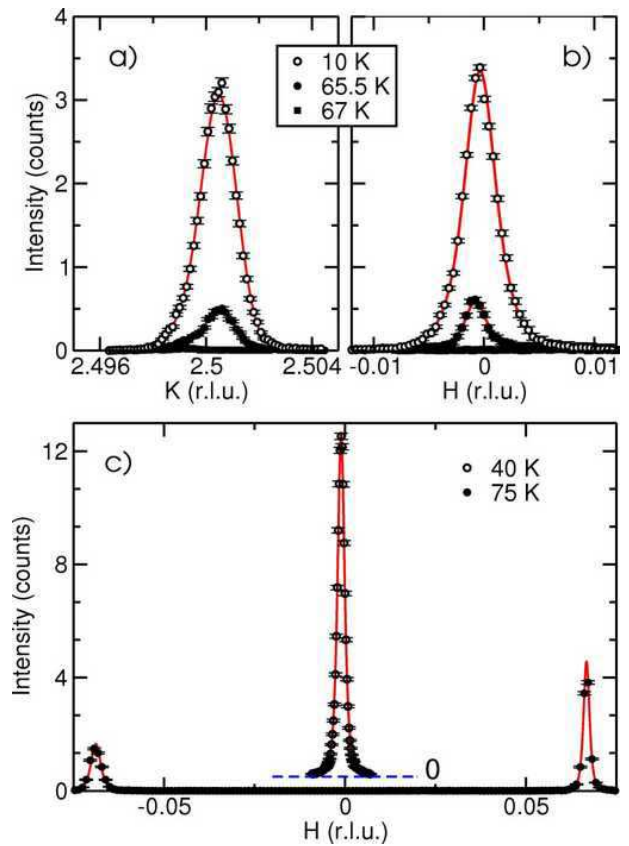}
  \caption{\label{fig:slxray} X-ray diffraction on a single crystal.  (a) Longitudinal
  scans and (b) transverse scans through the
    (0,~2.5,~0) commensurate position.  (c) Scans along
    $H$ through the incommensurate and commensurate peak positions.  The scans through
    the incommensurate peaks were performed at a different value of $K$ than the
    commensurate peak.}
\end{figure}

\begin{figure}[t!]
  \includegraphics[width=0.45\textwidth]{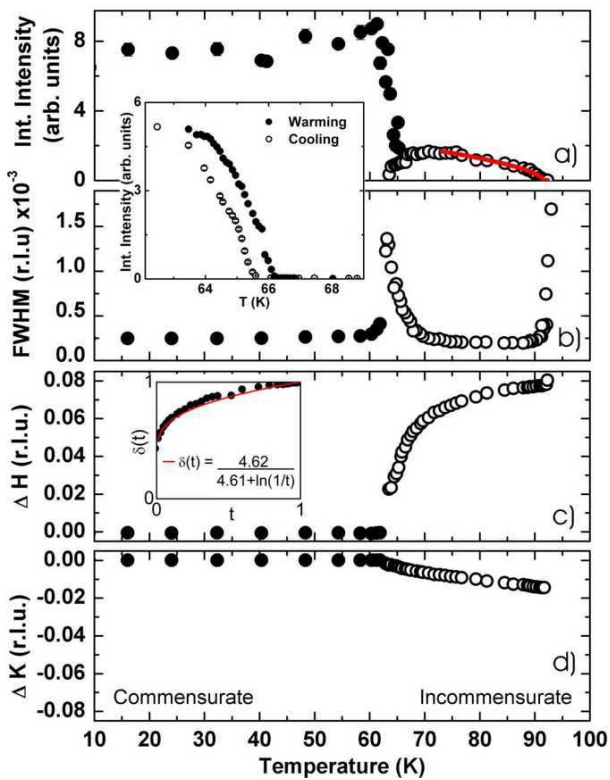}
  \caption{\label{fig:incommensurate} Temperature dependence of several fitted
    parameters for the commensurate and incommensurate superlattice peaks.  a) The
    integrated intensity, where the closed symbols refer to a commensurate
    peak and the open symbols refer to an incommensurate peak.
    b) The full-width at half-maximum (FWHM) along the $H$-direction.
    Inset: Hysteresis in temperature of the
    commensurate peak intensity.  c) and d) The components $\Delta H$ and $\Delta K$ of the
    displacement wave vector from the commensurate position.  Inset: Normalized
    displacement wave vector for the incommensurate modulation compared with a theoretical prediction
    involving discommensurations.}
\end{figure}

Synchrotron x-ray scattering measurements were performed on our
single crystal samples using the X20A and X22C beamlines at the
National Synchrotron Light Source at Brookhaven National Laboratory.
The samples were cooled using closed-cycle $^4$He refrigerators.
Upon cooling below $T_{c1}=66$ K, superlattice peaks were observed
at positions displaced from the fundamental Bragg peaks by
commensurate wavevectors (0,~$\pm \frac{1}{2}$,~0).  This indicates
doubling of the unit cell along the $b$-direction, consistent with a
dimerization of the lattice.  Scans along $H$ and $K$ through the
(0,~2.5,~0) superlattice peak (shown in Fig.~\ref{fig:slxray}(a) and
(b)) are resolution-limited, indicating that the dimerized structure
has long-range order. The integrated intensities of 45 fundamental
superlattice peaks were measured and used to determine the dimerized
structure. The intensities were fit using a model with two
adjustable parameters: $\delta$ (the Ti ion displacement along the
$b$-direction) and $\tau$ (the relative shift of neighboring Ti
chains along the $b$-direction).  Note that once the chains are
dimerized, there is no symmetry preventing a relative shift $\tau$
between the two chain sublattices.  The best fit values are
$\delta=0.03(1)b$, and $\tau=0.05(2)b$.  For simplicity, our model
neglected displacements of the Cl and O atoms, yet we still find
good agreement with the refinement parameters of Shaz {\em et al.}
for the dimerized state \cite{Shaz05}.

The temperature dependence of the integrated intensity of the
(0,~2.5,~0) superlattice peak is shown in
Fig.~\ref{fig:incommensurate}(a) as the closed circles.  Upon
warming, the intensity abruptly drops at $T_{c1}=66$~K, the same
temperature at which our thermodynamic measurements indicate a first
order transition.  In fact, the intensity of the commensurate
superlattice peak also exhibits a thermal hysteresis, shown in the
inset, with a width for the hysteresis loop similar to that measured
for the susceptibility. Therefore, both magnetic and structural
probes indicate the first order nature of transition at $T_{c1}$.

Upon warming above $T_{c1}=66$~K, the structure has an
incommensurate
modulation\cite{Imai03,Abel04,Rueckamp05,Krimmel06,Schoenleber06},
as shown in Fig.~\ref{fig:slxray}(c). That is, between $T_{c1}$ and
$T_{c2}=92$~K, superlattice peaks are found at positions displaced
from the fundamental positions by wave vectors ($\pm \Delta
H,~\frac{1}{2}\pm \Delta K,~0$). Here, the incommensurate wave
vectors are shifted from the commensurate wave vector
($0,~\frac{1}{2},~0$) by the displacement vectors ($\pm \Delta H,
\pm \Delta K, 0$).  The widths along the $H$-direction of both the
commensurate and incommensurate peaks are plotted in
Fig.~\ref{fig:incommensurate}(b).  Both sets of peaks are found to
be resolution-limited, indicating long correlation lengths (greater
than 2000~\AA), except for narrow temperature ranges in the
vicinities of $T_{c1}$ and $T_{c2}$. Along the $K$-direction, the
peaks remain resolution-limited over the measured temperature range.
The incommensurate wave vectors continuously change as a function of
temperature, and upon cooling approach the commensurate wave vector.
The values for $\Delta H$ and $\Delta K$ for an incommensurate peak
displaced from (0,~2.5,~0) are plotted  as a function of temperature
in Figs.~\ref{fig:incommensurate}(c) and (d). The magnitude for
$\Delta H$ is about a factor of 5 larger than $\Delta K$ at
temperature immediately below $T_{c2}=92$~K. The temperature
dependence of the integrated intensity for the incommensurate peak
is shown in Fig.~\ref{fig:incommensurate}(a). The solid line denotes
a fit to a power law $(T-T_{c2})^{\beta}$ with $\beta=0.3$.  Unlike
the transition at $T_{c1}$, the transition at $T_{c2}$ appears
second order in nature.

\subsection{Model of the Incommensurate Structure}

Since the magnitudes of $\Delta H$ and $\Delta K$ for the
incommensurate phase are much smaller than $\frac{1}{2}$ (in
reciprocal lattice units), the structure may be thought of as an
incommensurate modulation of the dimerized structure.  In CuGeO$_3$,
the dimerized lattice of the ground state of becomes incommensurate
when a large enough magnetic field is applied.  The incommensurate
structure has been interpreted as a lattice of solitons which
proliferate along the chains, separating locally dimerized regions.
In CuGeO$_3$, the soliton lattice produces a modulation along the
chain direction only, hence the incommensurate wave vector has a
non-zero component along the chain direction only.  In TiOCl, the
incommensurate wave vector is displaced from the commensurate
position by two components: $\Delta H$ and $\Delta K$, hence, it is
slightly more complicated.

\begin{figure} [t!]
  \centering
  \includegraphics[height=0.65\textheight]{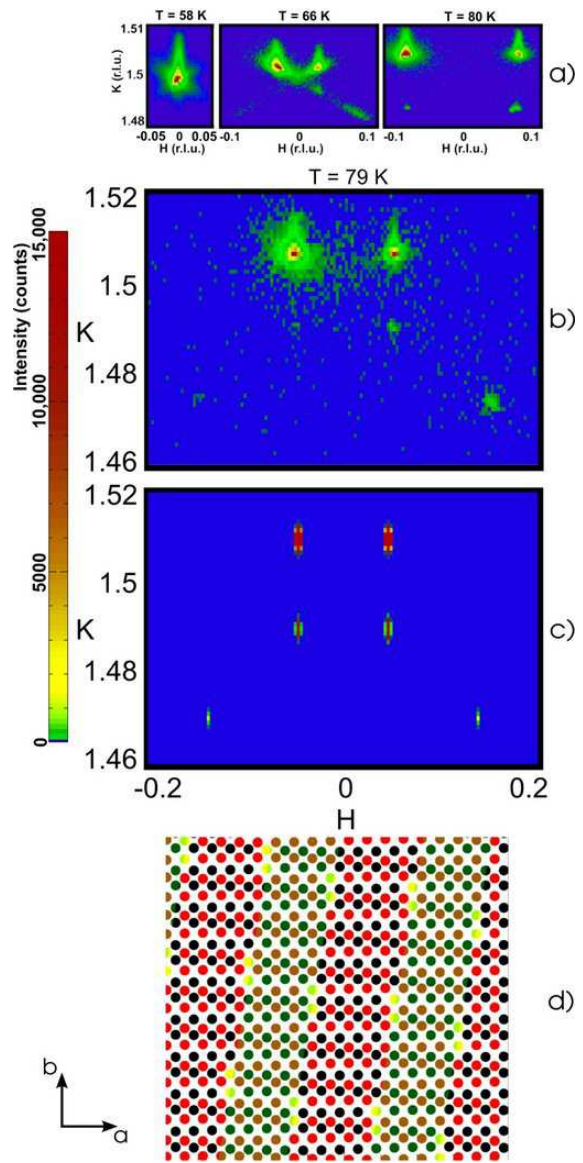}
  \caption{\label{fig:meshscans} (a) Series of intensity contour plots showing the temperature evolution
    of the superlattice peaks (b) Contour plot of the incommensurate
    peak data at $T=79$~K showing both the first and third harmonics.
    (c) Calculation of the intensity based on our model of the
    incommensurate structure, as discussed in the text.  (d)
    Real-space picture of the Ti atom positions in the incommensurate
    structure.  Alternate Ti chains have different colors (black and red) for easier viewing.
    The shaded (unshaded) domains delineate approximate regions with the same relative shift
    +$\tau$ ($-\tau$) between adjacent chains.
    The lightly colored atoms (yellow) on the domain walls denote solitons.}
\end{figure}

As an initial comparison, the temperature dependence of the
displacement wave vector magnitude $\Delta Q = \sqrt{\Delta H^2 +
\Delta K^2}$ is plotted in the inset of Fig.~\ref{fig:slxray}(c) and
compared to a Landau theory prediction.
Incommensurate-to-commensurate lock-in transitions have been
observed in charge density wave systems, such as
TaSe$_2$\cite{Moncton_CDW,Fleming}, upon cooling.
McMillan\cite{McMillan76} used a Landau expansion of the lattice
free energy to calculate the temperature dependence of $\Delta Q$
assuming that the deviations from the commensurate wave vector take
the form of long wavelength phase distortions or {\em
discommensurations}. The line in the inset of
Fig.~\ref{fig:slxray}(c) is the theory prediction\cite{McMillan76},
\begin{equation}
\label{eq:mcmillan} \delta\left(t\right)=4.62/[4.61+ln(1/t)],
\end{equation}
where $t=(T-T_{c1})/(T_{c2}-T_{c1})$ and $\delta(t) = \Delta
Q(t)/\Delta Q(t=1)$.  This prediction is a universal function of
$t$, and there are no adjustable parameters in the comparison.  The
theory roughly captures the singular behavior as $\delta \rightarrow
0$ near $T_{c1}$, suggesting that the incommensurate phase in TiOCl
may be described using a model of discommensurations separating
commensurate, dimerized regions. Intensity contour plots at
different temperatures are shown in Fig.~\ref{fig:meshscans}(a),
showing how the incommensurate peak positions converge to the
commensurate position as the temperature is cooled to $T_{c1}$.  The
observed broadening of the peaks just above $T_{c1}$ makes it
difficult to comment on possible phase coexistence\cite{Clancy} or a
discontinuous change in wave vector at the
incommensurate-to-commensurate transition.

To obtain a more specific picture of the incommensurate modulation,
we measured the intensities of several incommensurate superlattice
peaks in the vicinity of the (0,~1.5,~0) position.  At $T=79$~K, we
find four peaks near (0,~1.5,~0) displaced by wave vectors ($\pm
\Delta H,\pm \Delta K,0$) as depicted in the intensity map of
Fig.~\ref{fig:meshscans}(b).  All four of these peaks have the same
magnitude for the displacement vector, and we call these the first
harmonic peaks.  We were able to observe third harmonic peaks
displaced by ($\pm 3\Delta H,-3\Delta K,0$) from (0,~1.5,~0), also
shown in Fig.~\ref{fig:meshscans}(b).  This is the first time the
higher harmonic peaks have been observed, and their positions and
intensities provide crucial insight into the behavior of the
incommensurate modulation.  No signal above the background could be
observed at positions displaced by ($\pm 3\Delta H,+3\Delta K,0$)
indicating that the intensities of those peaks are very weak.  These
observations allow us to rule out a model consisting of the
superposition of two independent modulations along the $a$ and $b$
directions, as this would not yield third harmonic scattering at the
observed positions.

We analyze the intensities using a model which depends on the
real-space displacement vectors $\vec{u}_i$ of the atoms in TiOCl
from their equilibrium positions.  The resulting structure factor is
\begin{equation}
  \label{eq:incommsf}
  S(\vec{Q})=\sum_i f_i(Q) e^{\vec{Q}\cdot(\vec{R}_i+\vec{u}_i)}.
\end{equation}
where $f_i(Q)$ is the form factor for atom $i$ and the summation is
performed over an enlarged supercell for the modulated structure.
For simplicity, we only consider displacements of the Ti atoms,
leaving the O and Cl atoms at their equilibrium positions, and
relegate the displacements to be along the $b$-direction.

\begin{figure}[t!]
  \centering
  \includegraphics[height=.8\textwidth]{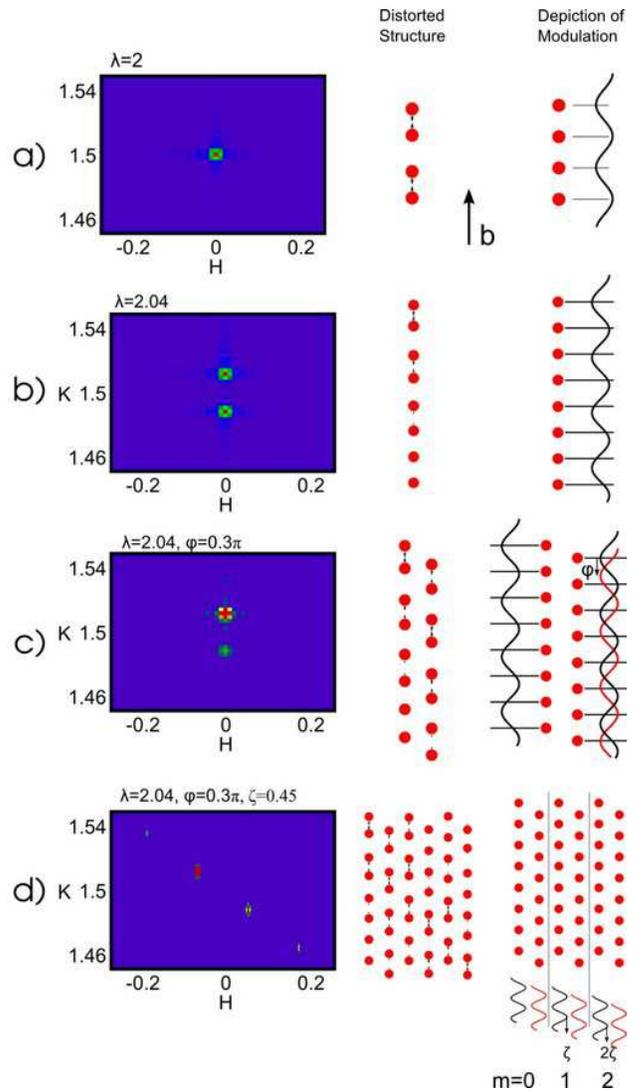}
  \caption{\label{fig:modulationexample}
  Real-space depictions of structural modulation along with corresponding reciprocal
  space maps of the structure factor.
  (a) Sinusoidal modulation for a single chain which yields a
  commensurate structure.
  (b) Incommensurate modulation obtained by changing the wavelength of the sinusoid.  (c)
  Modulation for two adjacent Ti chains, where the value for the relative phase
  shift $\phi$ yields an asymmetry in the intensities of the first harmonic peaks.
  (d) Model for the incommensurate modulation in TiOCl, obtained by phase shifting
  the modulation of each unit cell along the $a$-direction. }
\end{figure}

To construct the observed in pattern in Fig.~\ref{fig:meshscans}(b),
we begin by noting that the continuous evolution of $\Delta H$ and
$\Delta K$ with temperature indicates that the modulation is truly
incommensurate. Hence, we first consider the following sinusoidal
form for displacements along the $b$-direction for a single chain:
\begin{equation}
  \label{eq:sinmod}
  u(n)=\delta\cos\left(\frac{2\pi n}{\lambda}\right),
\end{equation}
where the integer $n$ refers to the $n$th Ti atom along $b$,
$\delta$ is the amplitude of the distortion, and $\lambda$ is the
wavelength of the modulation (in units of $b$).  The case of
$\lambda=2$ corresponds to a simple lattice dimerization and is
shown in Fig.~\ref{fig:modulationexample}(a) along with a reciprocal
space map of the corresponding commensurate superlattice peak.  An
incommensurately modulated structure can be obtained by making
$\lambda$ slightly larger or smaller than 2.
Figure~\ref{fig:modulationexample}(b) shows a modulation with
$\lambda=2.04$ along with the corresponding superlattice peak
pattern.

We observe significantly higher intensities for the first harmonic
peaks displaced by $+ \Delta K$ from (0,1.5,0) relative to those
displaced by $- \Delta K$ (Fig.~\ref{fig:meshscans}(b)).  This
naturally arises when one considers the two different Ti chains in
the unit cell.  By varying the relative phase shift $\phi$ between
the modulations on the two chains, different intensity ratios can be
obtained.  Figure~\ref{fig:modulationexample}(c) shows the
diffraction pattern corresponding to a unit cell with Ti chains
whose modulations differ by a relative phase shift $\phi = 0.3 \pi$.
The resulting structure factor reproduces the observed intensity
asymmetry.

Since the modulation is not simply along the $b$-direction (both
$\Delta H$ and $\Delta K$ are non-zero in the incommensurate phase),
we add an additional phase difference, $\xi$, between neighboring
pairs of Ti chains. The final expression for the displacements is
written as
\begin{equation}
  \label{eq:fullmod}
  u(n,m) = \delta\cos\left(\frac{2\pi n}{\lambda}+j \phi+\xi m\right)
\end{equation}
where $j=0,1$ labels the two different Ti chains within the unit
cell and the integer $m$ refers to the $m$th pair of Ti chains along
the $a$-direction.  A schematic for this modulation is shown in
Fig.~\ref{fig:modulationexample}(d). The addition of the phase shift
$\xi m$ has the effect of rotating the modulation direction. The
calculated third harmonic peak positions fall on the same line
connecting the first harmonic peaks on either side of (0,~1.5,~0),
as seen in the data.  This is {\em direct} evidence for a
one-dimensional (single wave) modulation describing the
incommensurate phase of TiOCl.  Now, there is a degeneracy between
displacements with opposite signs in front of $\phi$ and $\xi$. This
gives rise to two twin domains. By considering the other modulated
twin domain, the pattern in Fig.~\ref{fig:meshscans}(c) is obtained.
There is good agreement with the calculated intensity ratios in
Fig.~\ref{fig:meshscans}(c) and the observed ones in
Fig.~\ref{fig:meshscans}(b).  In addition, all of the peak positions
are reproduced.  We can also deduce the population factors of the
two twin domains to be about 60\% and 40\%.  In order to reproduce
the relative intensities of the third harmonics, we used a value for
$\delta$ of $0.2 b$, which is unrealistically large.  This indicates
that deviations from a pure sinusoidal modulation exist, as we
discuss below in the context of the higher harmonics peaks.

Figure \ref{fig:meshscans}(d) shows the real space positions of the
Ti atoms in our model for the modulation (in a single twin domain).
The adjacent staggered Ti chains are colored differently (red and
black) for easier viewing.  A local dimerization of the lattice is
readily seen throughout the plotted structure, where adjacent Ti
chains appear to have a small relative shift $\tau$ along the chain
direction.  For convenience of viewing, each shaded (unshaded)
region approximately delineates a single phase domain consisting of
the same relative shift $+\tau$ ($-\tau$). Note that shift $\tau$ is
not a parameter in the above model, but it naturally arises once
$\phi$ and $\xi$ are specified. On the boundaries between domains
(discommensurations), we find Ti atoms which are nearly undimerized
(colored yellow in Fig.~\ref{fig:meshscans}(d)).  These correspond
to locations where $u(n,m) \simeq 0$ and may be referred to as
solitons. Due to the sinusoidal form of the modulation in our
calculation, the solitons are not well defined, being spread out
over many Ti atoms. The other twin domain (with opposite signs of
$\phi$ and $\xi$) is similar to the structure in
Fig.~\ref{fig:meshscans}(d) but with domain walls which tilt in the
opposite sense with respect to the vertical chain direction.

\begin{figure}[t!]
  \includegraphics[width=0.45\textwidth]{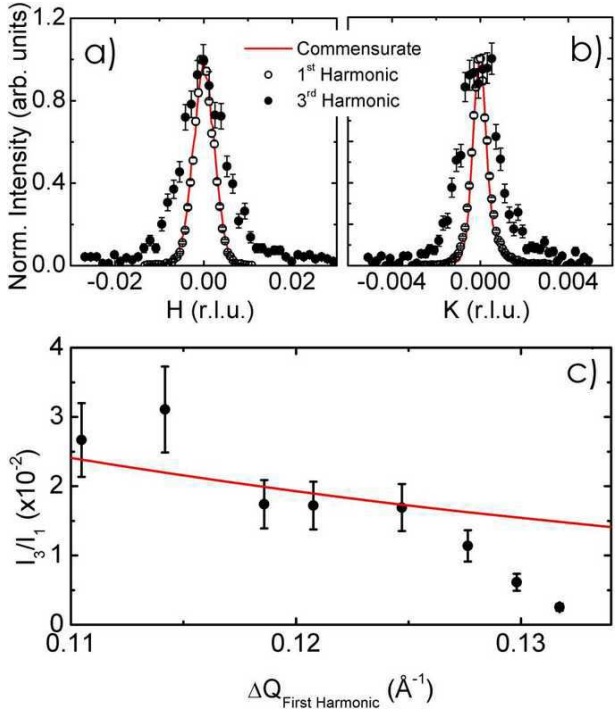}
  \caption{\label{fig:harmonics} (a) and (b) Scans through the commensurate, first harmonic incommensurate,
  and third harmonic incommensurate peaks along the $H$ and $K$ directions.  The
center of each peak was offset in order to plot the three scans on
the same axis.  (c) Ratio of the harmonic peak intensities
$I_3$/$I_1$ as a function of the displacement wave vector
    $\Delta Q$.  The line is the prediction from the soliton model as described in the text.}
\end{figure}

Figure~\ref{fig:harmonics}(a) shows scans along $H$ and $K$ through
the first and third harmonic peaks at $T=79$~K, along with
corresponding scans through the commensurate peak at $T=10$~K.  The
center of each peak was offset in order to plot the three scans on
the same axis.  The first harmonic peak and the commensurate peak
are resolution-limited, whereas the third harmonic peak shows a
clear broadening.  In CuGeO$_3$, Christianson {\em et
al.}~\cite{christianson02} observed a slight broadening of the
field-induced third harmonic peaks and attribute it to the effects
of disorder.  They concluded that the incommensurate phase was not
long-range ordered in the direction of the broadening, even though
it had a long enough correlation length to appear
resolution-limited.  We believe the same holds true for the
incommensurate phase in TiOCl in both the $a$ and $b$ directions.

The higher harmonic intensities can be calculated by expanding the
modulation in terms of the Fourier components; the ratio of harmonic
intensities is given by the ratio of the corresponding Fourier
coefficients.  Here, we follow an analysis similar to that used to
describe the field induced incommensurate modulation in CuGeO$_3$
\cite{kiryukhin96,kiryukhinkeimer96}. In TiOCl, we note that within
a single twin domain, the modulation may be considered to be
one-dimensional along the direction given by the displacement wave
vector $(\Delta H, \Delta K, 0)$.  When scattering from a periodic
one-dimensional modulation $u(x)$, the $i^{th}$ harmonic will have
intensity\cite{kiryukhin96}
\begin{equation} \label{eq:harmonicint}
I_n \sim \left(\int_0^\pi d\theta
\sin\left(i\theta\right)\sin\left[Q\;u\left(x\right)\right]\right)^2,
\end{equation}
where $\theta=qx$ and $q$ is the reduced wave vector along the
modulation direction. For the case of a sinusoidal function, we
calculate $I_3/I_1 \simeq 10^{-8}$ using a displacement amplitude of
$0.03 b$. The measured ratio at $T=79$~K is $I_3/I_1 \simeq 10^{-2}$
and, hence, is not consistent with a pure sinusoidal modulation. In
the limit of a square-wave function, we calculate an expected ratio
$I_3/I_1 \simeq 10^{-1} $, which is larger but closer to the
measured ratio.

For a better fit, we consider the Jacobi elliptic function
\begin{equation}
  \label{eq:icmod}
  u(l)=\epsilon\left(-1\right)^l sn\left(\frac{\lambda}{\Gamma
  k},k\right)
\end{equation}
where $l$ denotes a lattice site and the elliptic modulus $k$ can
have values between 0 and 1 ($k=0$ corresponds to a sine wave and
$k=1$ a square wave).  The parameter $\Gamma$ is the soliton half
width, and $\lambda/2$ is equal to the soliton spacing. Then, the
ratio $I_3/I_1$ can be written:
\begin{equation}
  \label{eq:i3i1}
  \frac{I_3}{I_1}=\left(\frac{Y}{Y^2+Y+1}\right)^2.
\end{equation}
where $Y=\exp\left[-\pi K \left(\sqrt{1-k^2}\right)/K(k)\right]$,
and $K$ is the complete elliptic integral of the first kind. The
soliton spacing, $\lambda/2$, can be written in terms of $k$ by
\begin{equation}
  \label{eq:lambdak}
 \frac{\lambda}{2}=\frac{\pi}{\Delta Q}=2 k K(k)\Gamma.
\end{equation}

At each temperature, the measured values for $I_3/I_1$ and the
magnitude of the displacement wave vector $\Delta Q$ can be used to
determine $k$ and $\Gamma$.  For the data in the range $\Delta Q <
0.125$, the calculated values for the soliton widths were roughly
constant with $\Gamma = 5(1)$~{\AA}, indicating relatively sharp
domain walls.  By keeping the soliton widths fixed at this value,
the expected ratio $I_3/I_1$ can be calculated as a function of
$\Delta Q$  as shown by the line in Fig.~\ref{fig:harmonics}(c). The
calculated line agrees reasonably well for $\Delta Q < 0.125$,
lending credence to the soliton lattice description for the
incommensurate phase of TiOCl.  In Fig.~\ref{fig:harmonics}(c), the
data points at larger $\Delta Q > 0.125$ have $I_3/I_1$ ratios which
are much smaller than the prediction.  This means that for these
temperatures, which are just below the onset temperature $T_{c2}$,
the soliton width is considerable larger than $\Gamma \simeq
5$~{\AA} and the modulation is more sinusoidal.  For the nearest
neighbor Heisenberg model, Nakano and Fukuyama predict
\cite{Nakano80} $\Gamma=\frac{\pi J b}{2 \Delta_o}$, where
$\Delta_o$ is the magnetic gap.  Using $J=660$~K\cite{Seidel03} and
$\Delta_o = 430$~K \cite{Imai03,Baker}, this yields
$\Gamma\simeq8$~\AA, which represents an approximate length scale
for the soliton width along the chains.  The experimental value
pertains to the soliton width along the incommensurate modulation,
which in TiOCl is not strictly along the chain direction.

Our results lead to the following picture for the sequence of
structural changes.  Just below $T_{c2}$, the structure has a
one-dimensional incommensurate modulation of an underlying dimerized
state.  At first, the envelope function for the modulation has a
relatively short wavelength and is approximately sinusoidal.  Upon
cooling, the wavelength increases and the modulation crosses over
from being sinusoidal to being better described by a soliton lattice
(with solitons of fixed width).  The solitons form domain wall
boundaries (or discommensurations) between locally dimerized domains
as shown in Fig.~\ref{fig:meshscans}(d). The domains on either side
of the discommensurations may be roughly characterized as having
opposite signs for $\tau$, the shift parameter between neighboring
chains. The soliton spacing (spacing between domain walls) increases
with decreasing temperature and diverges at $T_{c1}$. At this point,
the system wants to become uniformly dimerized, and a single $\tau$
must be selected. This selection between two degenerate ground
states with different symmetries may help explain the first order
nature of the transition at $T_{c1}$.\cite{Fausti} The above
description applies to each twin domain.

Our model, which focuses only on the Ti displacements, is consistent
with the structural refinements of Schonleber {\em et
al.}\cite{Schoenleber06}.  However, by measuring the higher harmonic
content, we have obtained additional details on the structure.  The
above picture also gives insight into the reason for the
incommensurate modulation.\cite{Rueckamp05,Schoenleber06} Assuming
the lattice energy prefers an equal spacing between Ti atoms between
chains, then for uniformly dimerized chains, this preference cannot
be satisfied due to the staggered nature of the chains. The
sinusoidal modulation allows the lattice energy cost to be spread
out over many unit cells. As the dimerization amplitude increases,
it becomes energetically favorable to form fully dimerized domains
at the local level, with the cost in lattice energy being relegated
to domain walls.  Eventually upon cooling, the magnetic energy gain
due to dimerization dominates, and the lattice prefers a uniformly
dimerized state with the value for $\tau$ which minimizes the
lattice energy cost.

\section{Inelastic X-ray Scattering}
\subsection{\label{sec:inelastic}Lattice Dynamics}

The lattice dimerization which accompanies the spin-Peierls
transition is expected to be a displacive structural phase
transition.  Such displacive transitions are characterized by a
softening of a zone-boundary phonon mode which has a polarization
similar to that of the static
distortion.\cite{Shirane70,Shapiro72,Shirane74} This mode softens
progressively as the temperature is decreased, until the mode energy
reaches zero, and the distortion becomes frozen in.  In their theory
for the spin-Peierls transition, Cross and Fisher have calculated
the expected behavior of such a soft mode.\cite{CrossFisher79}
However, to date, the lattice dynamics of a zone-boundary soft
phonon in a spin-Peierls system have not been studied.  Here, we
present a detailed inelastic x-ray scattering characterization of
the lattice dynamics in TiOCl. Thus far the only work on the lattice
dynamics of TiOCl has been Raman and infrared spectroscopy
measurements of the zone center optical modes
\cite{lemmens03,caimi04,Fausti}. The inelastic x-ray scattering
technique has the distinct advantage in being able to probe the
phonon modes throughout the Brillouin zone. Also, while inelastic
neutron scattering requires relatively large crystals, inelastic
x-ray scattering can measure very small crystals of interesting new
correlated electron systems.

\begin{figure}[t!]
  \centering
  \includegraphics[width=.45\textwidth]{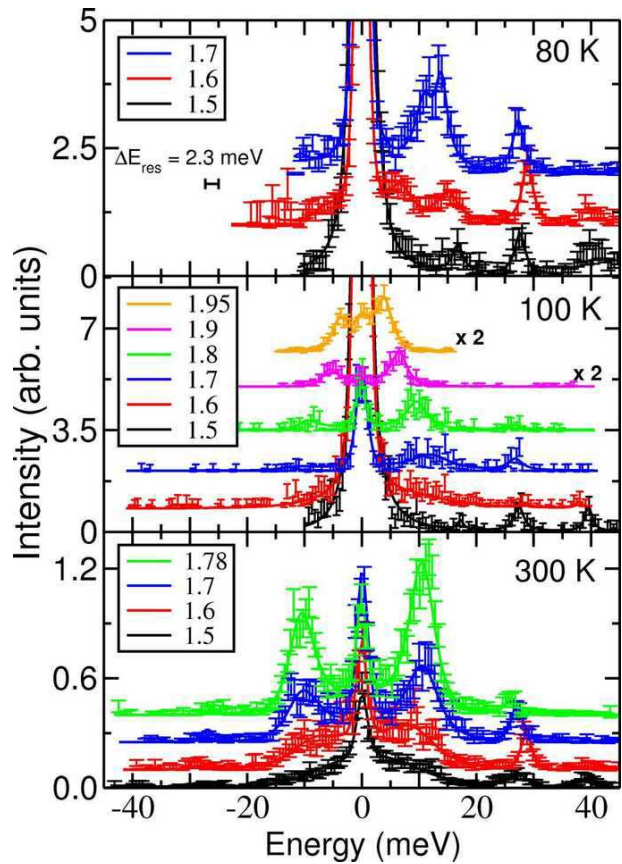}
  \caption{Energy transfer scans through the phonon excitations at different $\vec{Q}$
  positions along (0 $K$ 0), at temperatures of (a) $T=80$~K, (b) $T=100$~K, and (c) $T=300$~K.
    The solid lines are fits to a damped harmonic oscillator structure factor convoluted with the instrumental
    resolution as described in the text.}
  \label{fig:phonons}
\end{figure}

The inelastic x-ray scattering measurements were performed using the
SRI-3ID beamline at the Advanced Photon Source at Argonne National
Laboratory.  An in-line nested channel-cut silicon monochromator and
a bent silicon analyzer in backscattering geometry were used to
achieve high energy resolution.  By scanning the energy transfer
$\omega$ through the elastic scattering from a plexiglass sample,
the energy resolution can be determined. The instrumental resolution
function has the following approximate form
\begin{align}
  \label{eq:sector3res}
  I\left(\omega\right)=I_o\left\{\frac{2\eta}{\pi\gamma}\left[1+4\left(\frac{\omega}
        {\gamma}\right)^2\right]^{-1}+ \left(1-\eta\right)\frac{2}{\gamma}
    \left(\frac{\ln{2}}{\pi}\right)^{1/2}\right.\nonumber\\
  \times \left.\exp{\left[-4 \ln{2}\left(\frac{\omega}{\gamma}\right)^2\right]}\right\},\nonumber\\
\end{align}
where $\gamma$ is the full-width and half-maximum (FWHM), $I_o$ is
the integrated intensity, and $\eta$ is a mixing parameter
\cite{Sinn01}.  We determined the spectrometer energy resolution to
be 2.3 meV FWHM.  The data we present below have been corrected for
temperature drifts of the monochromator and are normalized by
monitor counts.

\begin{figure}[t!]
  \centering
  \includegraphics[width=.4\textwidth]{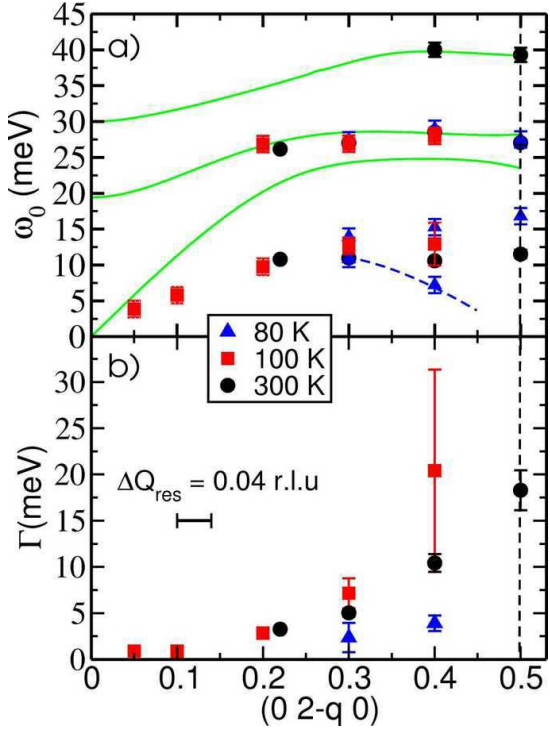}
  \caption{\label{fig:dispersion} (a) Dispersion relations for the observed phonon modes where
  the undamped frequency $\omega_o$, extracted from the
  damped harmonic oscillator fits, is plotted.  The lines are predictions from a lattice
  dynamical calculation. (b) The damping
parameter $\Gamma$ for the lowest energy mode.  }
\end{figure}

We performed experiments on two samples of TiOCl, consisting of
crystals coaligned to within 2-3 degrees with total thicknesses of
$\sim 20$ $\mu$m and $\sim 40$ $\mu$m.  The incident photon energy
was 21.3 keV.  Representative energy scans are shown in
Fig.~\ref{fig:phonons} at temperatures of 80~K, 100~K, and 300~K.
For each scan, $\vec{Q}$ was held constant at various points along
(0,~$K$,~0) between (0,~1.5,~0) and (0,~2,~0).  The scattered
intensity is a direct measure of the dynamic structure factor
$S(\vec{Q},\omega)$ which is related to the dissipative part of the
response function by the fluctuation-dissipation theorem.  To model
the lattice dynamics, we used the damped harmonic oscillator
response function:
\begin{equation}
  \label{eq:dampedoscillator}
  \centering
  S(\vec{Q},\omega)=\frac{A_{\vec{Q}}}{\pi}\frac{2\omega\Gamma}{(\omega_o^2-\omega^2)^2+\omega^2\Gamma^2}
  \left[n(\omega)+1\right],
\end{equation}
where $\omega_o$ is the undamped phonon frequency, $\Gamma$ is the
damping constant, and $A_{\vec{Q}}$ is an amplitude.  In the
underdamped case ($\omega_o > \Gamma$) two inelastic peaks are
present at positive and negative energy transfers.  In the
overdamped case ($\omega_o < \Gamma$) the scattering has the form of
a single peak centered about $\omega=0$. In the extreme overdamped
limit ($\omega_o \ll \Gamma$) the dynamic structure factor can be
rewritten as
\begin{equation}
  \label{eq:overdamped}
  S(\vec{Q},\omega) \approx \frac{k_B T}{\hbar \omega_o^2} \frac{1}{\pi}\frac{\gamma}{\gamma^2 + \omega^2},
\end{equation}
where $\gamma=\omega_o^2/\Gamma$ and the high temperature
approximation of $1+n(\omega) \simeq k_B T/\hbar \omega$ has been
used.  The solid lines in Fig.~\ref{fig:phonons} are fits to the
damped harmonic oscillator cross section convoluted with the
instrumental resolution function. In some scans, up to three phonon
modes can be fit.

\begin{table}[t!]
  \includegraphics[width=.49\textwidth]{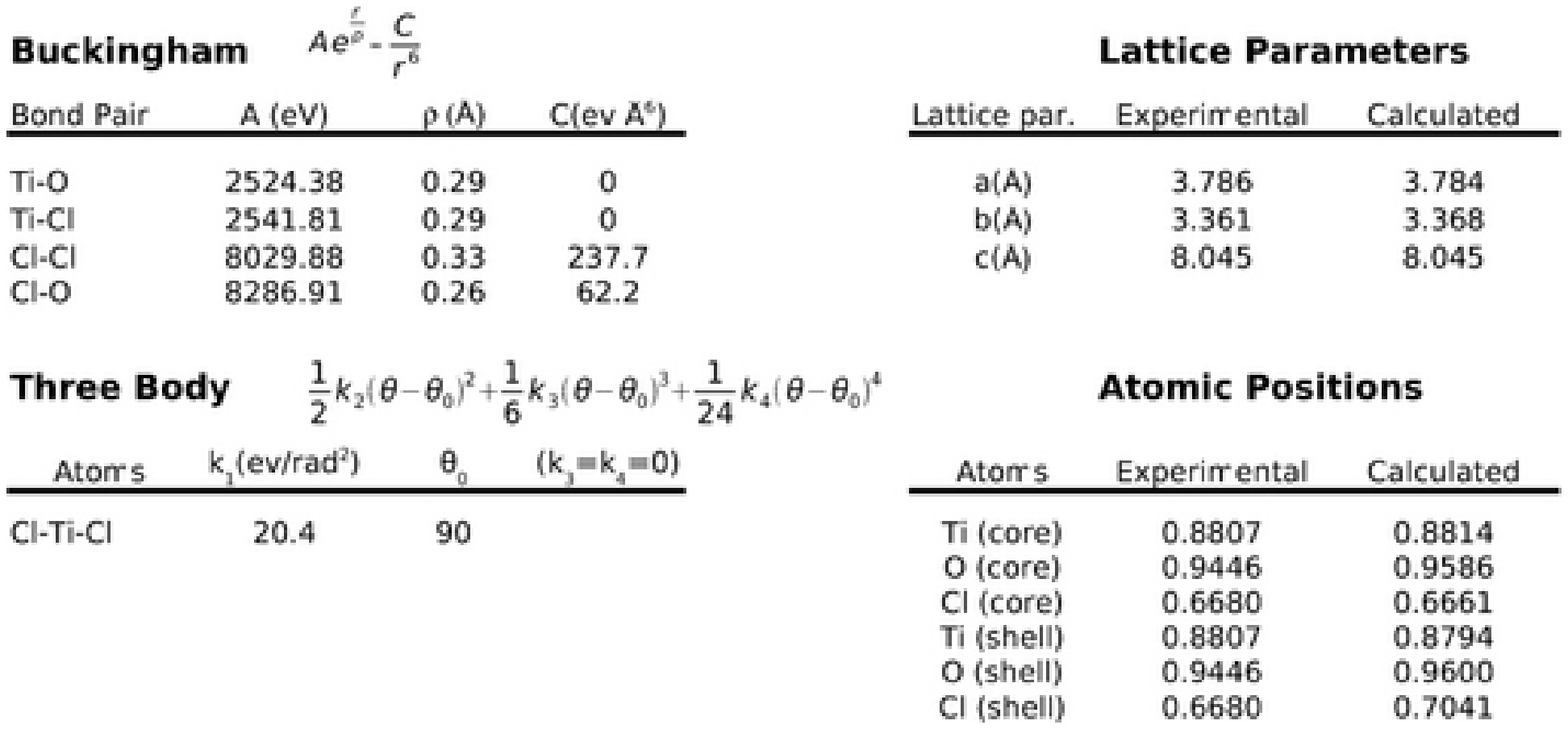}
  \caption{\label{tab:shellparameters}
    Interaction parameters used in the shell model calculation, and the resulting equilibrium
    crystal parameters.}
\end{table}

Figure \ref{fig:dispersion}(a) shows the fitted values for
$\omega_o$ for all of the observed phonons as a function of the
reduced momentum transfer along the (0 $K$ 0) direction. Note that
in this scattering geometry, the cross section is such that only
longitudinally polarized phonons are measured. The lowest energy
mode corresponds to the longitudinal acoustic branch.  The widths of
the phonons in this branch are plotted in
Fig.~\ref{fig:dispersion}(b).  At $T=300$~K (well above
$T_{c2}=92$~K), the width of the phonon mode increases as $\vec{Q}$
approaches the zone-boundary.  In fact, at the zone-boundary, the
fit yields $\Gamma > \omega_0$ which indicates that the phonon is
overdamped.  The data at $T=100$~K (near $T_{c2}$) show even more
dramatic behavior.  The width of the phonon mode diverges well
before the zone-boundary is reached, indicating that the mode is
strongly overdamped.  These results are the first indication that
the longitudinal acoustic phonon at the zone-boundary is a soft
mode.

\begin{figure}[t!]
  \centering
  \includegraphics[width=.45\textwidth]{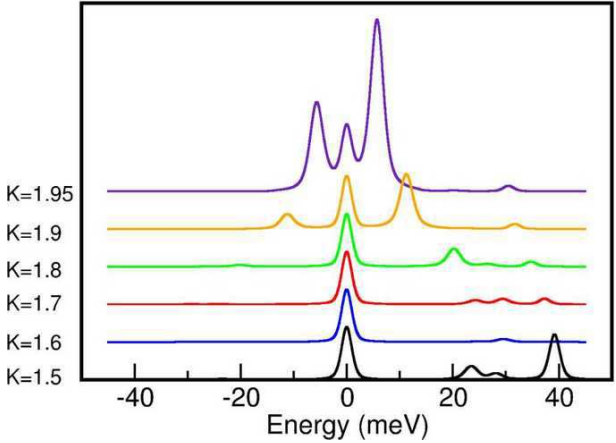}
  \caption{\label{fig:calcsf}
   Calculated cross sections for the longitudinal phonons in the undimerized state at
   various (0 $K$ 0) positions, using
   the eigenvectors and eigenvalues from the lattice dynamical calculation.}
\end{figure}

\begin{figure}[b!]
  \centering\vspace{8mm}
  \includegraphics[width=.45\textwidth]{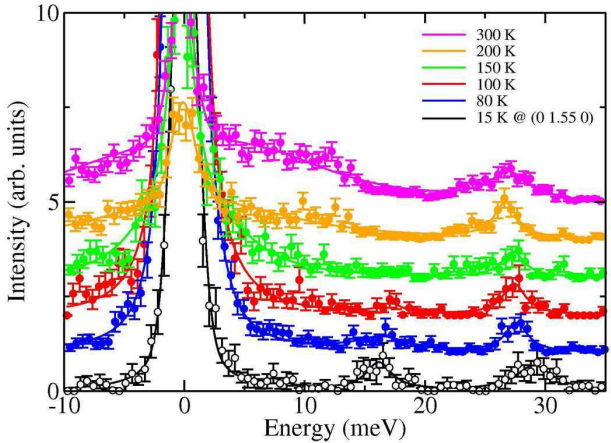}
  \caption{Energy transfer scans at (0,~1.5,~0) for different temperatures.
    The data have been offset for clarity. The $T = 15$~K data were
    measured at (0,~1.55,~0) to avoid contamination from the elastic superlattice peak at $\omega=0$.}
  \label{fig:phonontdep}
\end{figure}

The solid lines in Fig.~\ref{fig:dispersion}(a) correspond to
lattice dynamical calculations for the longitudinal phonon modes
based on a shell model using the General Utility Lattice
Program~\cite{GULP}.  The calculation uses phenomenological
potentials to model the strong short-range interactions of the
neighboring ions\cite{gulpweb}.  Four Buckingham potential
interactions serve as a repulsive force between atoms, and an
additional three-body interaction was used to constrain an angle of
$90^\circ$ for the Ti$-$Cl$-$Ti bonding.  With these potentials, the
system settled into an equilibrium structure nearly identical to the
experimental unit cell~\cite{Schaefer58} (see Table
\ref{tab:shellparameters}). A comparison between the calculated and
observed optic phonons in Fig.~\ref{fig:dispersion}(a) shows
excellent agreement.  However, the calculation significantly
overestimates the energy for the acoustic branch near the zone
boundary, another indication of the soft mode.  We note that our
calculation and others\cite{Yildirim} indicate that this mode is
degenerate at the zone boundary.  The energy eigenvalues and
eigenvectors resulting from the lattice dynamical calculations allow
us to calculate the expected scattering intensities using the
dynamical structure factor convoluted with the instrumental
resolution.  This calculation is shown in Fig.~\ref{fig:calcsf}. The
agreement with the observed intensities that we measure is quite
good.  Again, deviations from the data are apparent for low energies
near the zone-boundary due to the presence of the soft mode.

\begin{figure}[b!]
  \centering\vspace{8mm}
  \includegraphics[width=.4\textwidth]{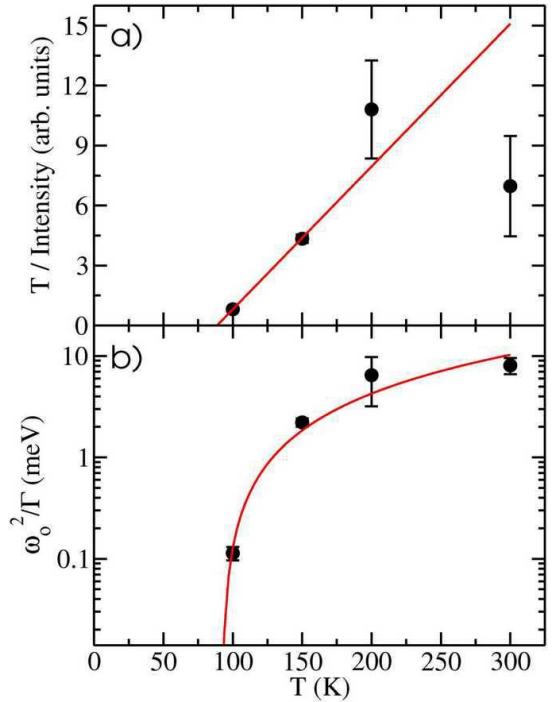}
  \caption{Behavior of the soft phonon at the zone boundary position (0,~1.5,~0).
  (a) The quantity $T$/Intensity as a function of temperature.  The solid
line follows the form $a(T-T_c)$ with $T_c =88$~K. (b) The quantity
$\omega_o^2/\Gamma$ as a function of temperature. The solid line
serves as a guide to the eye.}
  \label{fig:overdampedtdep}
\end{figure}

The temperature dependent behavior of the zone boundary phonon is
shown in  Fig.~\ref{fig:phonontdep}.  The data for $T=300$~K and
$T=200$~K were fit to Eq.~\ref{eq:dampedoscillator} (the general
cross section for the damped harmonic oscillator).  However at
$T=150$~K and below, the fit no longer converged, and
Eq.~\ref{eq:overdamped} was used for the strongly overdamped case.
In this case, it is no longer possible to determine the values of
$\omega_o$ and $\Gamma$ independently.  Useful information can be
extracted by noting that the integrated intensity $I$ is directly
proportional to $T/\omega_o^2$.  Therefore, a plot of $T/I$ versus
$T$ (shown in Fig.~\ref{fig:overdampedtdep}(a)) should have the same
temperature dependence as $\omega_o^2$. The solid line has the form
$\omega_o^2 = a(T-T_o)$ which is the result from mean-field theory
for a soft-phonon transition,\cite{Shapiro72,Shirane74} where $T_o$
is the structural transition temperature and $a$ is a constant. The
fitted line yields $T_o = 88(2)$~K which closely matches the
temperature $T_{c2} = 92$~K for the onset of the incommensurate
superlattice peaks. The quantity $\omega_o^2/\Gamma$ can also be
extracted and is plotted in Fig.~\ref{fig:overdampedtdep}(b).  We
see that this quantity plunges to zero around 90~K.  This is
consistent with $\omega_o \rightarrow 0$ (accompanied by an increase
in $\Gamma$) as expected for a soft phonon transition. Therefore,
our inelastic x-ray scattering data provide clear evidence for a
soft phonon transition occurring in TiOCl. Specifically, we find
that a softening of the longitudinal acoustic phonon at the
zone-boundary drives the structural transition to the incommensurate
(nearly dimerized) state at $T_{c2} = 92$~K.

\subsection{Comparison with the Cross-Fisher Theory}
\label{sec:tioclspinphonon}

The preceding analysis relied on fits to a phenomenological damped
harmonic oscillator model.  While this provides strong evidence for
the presence of a soft mode, details regarding the microscopic
interactions, such as the spin-phonon coupling, cannot be obtained.
A general response function which includes such interactions has the
form
\begin{equation}
  \label{eq:A}
  A(\omega) = \frac{1}{\pi}Im\left[\frac{1}{\omega^2-\Omega_0^2-\Pi(\vec{Q},\omega)}\right],
\end{equation}
where $\Omega_0$ is the harmonic or bare phonon frequency, and
$\Pi(\vec{Q},\omega)$ is the polarizability (or phonon-self energy).
The dynamic structure factor is then given by
\begin{equation}
  \label{eq:Smu}
  S(\vec{Q},\omega)=|S(\vec{Q})|^2\left[1+n(\omega)\right]A(\omega).
\end{equation}
By comparing this cross section with that for the damped harmonic
oscillator, one can see that $\Pi(\vec{Q},\omega)$ contains
information regarding the damping. Since $\Pi(\vec{Q},\omega)$ is
the only term which is complex, it can be broken into real and
imaginary parts
\begin{equation}
  A(\omega) = \frac{1}{\pi}\frac{Im\left[\Pi(\vec{Q},\omega)\right]}{(\omega^2-\Omega_0^2 -
  Re\left[\Pi(\vec{Q},\omega)\right])^2+Im\left[\Pi(\vec{Q},\omega)\right]^2}.
\end{equation}
If we then make the substitutions
\begin{align}
  \label{eq:PiSubs}
  Re\left[\Pi(\vec{Q},\omega)\right] & = \omega_o^2(\vec{Q},\omega)-\Omega_0^2(\vec{Q})\\
  Im\left[\Pi(\vec{Q},\omega)\right] & = \omega\Gamma(\vec{Q},\omega),
\end{align}
we recover the damped harmonic oscillator response function, where
$\omega_0=\sqrt{\Omega_0^2+Re[\Pi]}$.  Here, $\omega_0$ represents
the ``quasi-harmonic'' frequency which differs from the bare phonon
frequency due to the presence of spin-phonon coupling or other
anharmonic interactions.

Cross and Fisher (CF) calculated the polarizability
$\Pi(\vec{Q},\omega)$ which takes into account the spin-phonon
coupling. The phonon system was treated using a mean-field random
phase approximation (RPA), whereas the spin dynamics were calculated
in a non-perturbative fashion, more accurate that the Hartree
approach. Their treatment begins by considering a set of
non-interacting spin chains, where each chain is governed by the
nearest neighbor Heisenberg Hamiltonian:
\begin{equation}
  \label{eq:AFH}
  H_s=\sum_l J(l,l+1)\mathbf{S}_l\cdot\mathbf{S}_{l+1}.
\end{equation}
Here, the planes of atoms perpendicular to the chains are
constrained to move together.  Assuming a linear dependence of $J$
on lattice distortions, the exchange coupling may be written
\begin{equation}
  \label{eq:CFJ}
  J(l,l+1) = J+\frac{1}{\sqrt{N}}\sum_q g(q)Q(q)e^{i q b}\left(1-e^{iq l b}\right),
\end{equation}
where $q$ is the reduced wave vector along the chain direction,
$g(q)$ is the spin-phonon coupling, and $Q(q)$ denotes the phonon
normal mode coordinates. The spin-phonon interaction leads to an
expression for the polarizability that depends on the dimer-dimer
correlation function
$\langle\left[\left(\mathbf{S}_l\cdot\mathbf{S}_{l+1}\right)_{t},
\left(\mathbf{S}_0\cdot\mathbf{S}_1\right)_{t=0}\right]\rangle$
which CF calculate using bosonization.  It is also possible to
calculate the dimer-dimer correlation function using conformal field
theory~\cite{Pouyan07}, which yields an identical result in the long
wavelength limit. The CF polarizability is calculated to
be~\cite{CrossFisher79}
\begin{equation}
  \label{eq:CFPi}
  \begin{split}
  \Pi_{CF}(q,&\omega) = -0.37\;|(1-e^{iqb})g(q)|^2\\
  &\times I_1\left(\frac{\omega+c(q-2k_f)}{2\pi T}\right) I_1\left(\frac{\omega-c(q-2k_f)}{2\pi
  T}\right)\frac{1}{T},
  \end{split}
\end{equation}
where
\begin{equation}
  \label{eq:CFI1}
  I_1(k) = \frac{1}{\sqrt{8\pi}}\frac{\Gamma\left(\frac{1}{4}+\frac{1}{2}ik\right)}
{\Gamma\left(\frac{3}{4}+\frac{1}{2}ik\right)},
\end{equation}
where $2k_f=\pi/b$ and $c=\pi J b/2$~\cite{Cloiseaux} is the
spin-wave velocity. In this treatment, the condition that defines
the spin-Peierls phase transition temperature, $T_{SP}$, is
\begin{equation}
  \label{eq:CFcriteria}
  \omega_o^2 = \Omega_0^2+\Pi_{CF}(q=2k_F,\omega\rightarrow 0,T=T_{SP}) = 0.
\end{equation}
This can be solved to get an expression for the spin-phonon coupling
for the zone boundary mode:
\begin{equation}
  \label{eq:CFg}
  g=\Omega_0\sqrt{\frac{\pi}{3.2} T_{SP}}.
\end{equation}

\begin{figure}[t!]
  \centering
  \includegraphics[width=.45\textwidth]{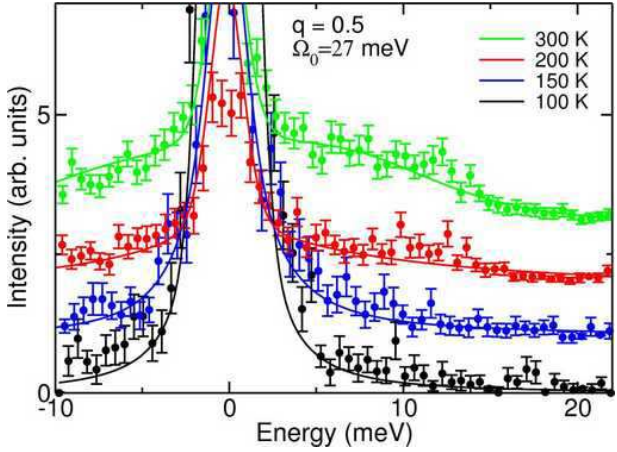}
  \caption{\label{fig:cftdepraw}
    Energy transfer scans of the low energy phonon mode at the zone boundary.  The
    lines denote fits to the dynamical structure factor calculated using the Cross-Fisher
    theory.}
\end{figure}

We now compare our data with the results of Cross and Fisher by
fitting our data with the dynamic structure factor that explicitly
includes the CF polarizability $\Pi_{CF}(q,\omega)$
(Eq.~\ref{eq:CFPi}) convoluted with the instrumental resolution.
Only data in the energy range -20~meV$\le \omega \le 20$~meV was
fit, thereby focusing on the longitudinal acoustic mode.  We first
fit data taken the zone boundary position (0,1.5,0) at several
temperatures, as shown in Fig.~\ref{fig:cftdepraw}. At the zone
boundary, the CF polarizability reduces to
\begin{equation}
  \label{eq:zbCF}
  \Pi_{CF}(q,\omega)=-0.74g^2 \left[I_1\left(\frac{\omega}{2\pi
  T}\right)\right]^2\frac{1}{T}.
\end{equation}
Therefore, the only adjustable parameters in the dynamical structure
factor are the spin-phonon coupling $g$ and the bare phonon
frequency $\Omega_0$. Since $g^2$ scales with the CF polarizability,
a non-zero $g$ will affect both the peak width and the shift of
$\omega_0$ from $\Omega_0$. Away from the zone boundary, the
polarizability also depends on $J$, therefore $J$ and $g$ cannot be
fit independently in general.

The solid lines in Fig.~\ref{fig:cftdepraw} represent fits to the
dynamical structure factor at $q=0.5$ based on the CF theory.  Here,
we let $g$ vary with temperature, but kept $\Omega_0$ fixed at the
best fit value of $\Omega_0=27(1)$~meV.  Letting $g$ float with
temperature was necessary in order to get reasonable fits, and we
will comment on this later.  Once $g$ was determined for a given
temperature, the scans at other $q$ positions were then fit with
$\Omega_0$ and $J$ as the adjustable parameters.  For the data sets
at $T=100$~K and $T=300$~K, the analysis yielded a roughly
consistent magnetic exchange of $J \sim 200$~K.  This value for $J$
is about three times smaller than that deduced from the magnetic
susceptibility results\cite{Seidel03}.  However, our value for $J$
is remarkably close considering that it resulted from measurements
of only the {\em phonon} positions and lineshapes.

\begin{figure}[t!]
  \centering
  \includegraphics[width=.45\textwidth]{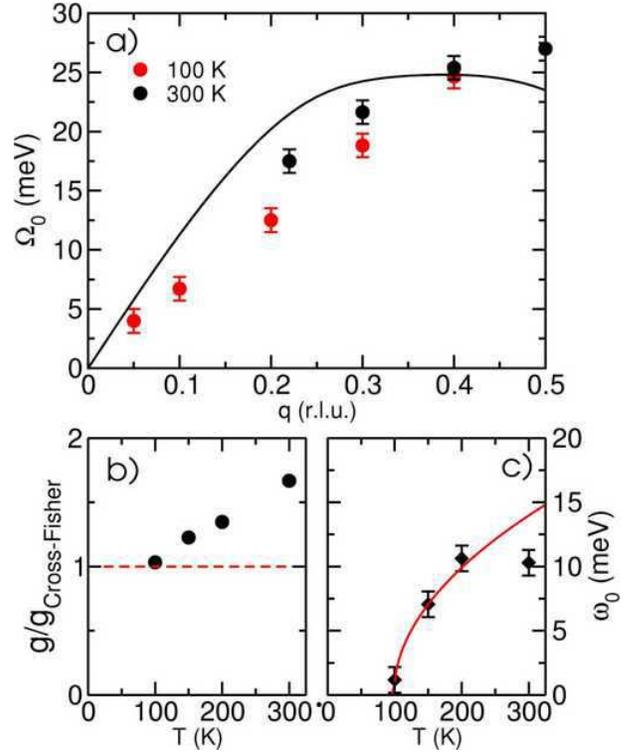}
  \caption{\label{fig:cffitsummary} (a) The bare phonon frequency $\Omega_0$ of
    the longitudinal acoustic mode extracted from the fits to the Cross-Fisher theory.
    The solid line denotes the shell model calculation.
    (b) The fitted spin-phonon coupling $g$, normalized to the value of $g$
    calculated using Eq.~\ref{eq:CFg}.
    c) The quasi-harmonic frequency $\omega_o$ of the zone boundary mode, extracted from fits to the
    Cross-Fisher theory.  The solid line denotes the power-law form $A\sqrt{T-T_c}$ with $T_c=98$~K.}
\end{figure}

The resulting values for the bare phonon frequency $\Omega_0$ are
plotted in Fig.~\ref{fig:cffitsummary}(a) as a function of the
reduced wave vector $(0,q,0)$. The solid line denotes the prediction
from the lattice dynamical calculation discussed in the previous
section. Clearly, the bare phonon frequency $\Omega_0(q)$ follows
the calculated dispersion curve much better than the mode energies
$\omega_o$ from the damped harmonic oscillator fits
(Fig.~\ref{fig:overdampedtdep}(a)).  Thus, it appears that the CF
theory captures the anharmonic effects of the spin-phonon coupling
in TiOCl reasonably well. However, in order to obtain good fits at
all temperatures, $g$ was allowed to vary. These values are plotted
in Fig.~\ref{fig:cffitsummary}(b), normalized by the value of $g$
calculated using Eq.~\ref{eq:CFg} (with $\Omega_0=27$~meV and
$T_{c2}=92$~K), which yields $g_{Cross-Fisher}=75$ meV$^{3/2}$.  The
figure shows that the fitted values of $g$ approach
$g_{Cross-Fisher}$ as $T \rightarrow T_{c2}$. Since the fits to the
dynamical structure factor were performed without knowledge of
$T_{c2}$, this result supports the self-consistency and validity of
the CF expressions describing the soft phonon near $T_{SP}$.  We
also have enough information to determine the quasi-harmonic
frequency $\omega_o$ at $q=0.5$, which is plotted in
Fig.~\ref{fig:cffitsummary}(c).  We see that the spin-phonon
coupling reduces $\omega_o$ from the bare value $\Omega_0$ by about
a factor of 2.  Also, at the spin-Peierls transition, $\omega_o$
should fall to zero. Indeed, the solid line through the data is a
fit to the form $A\sqrt{T-T_c}$ where $T_c=98$~K, which is close to
$T_{c2}=92$~K. Hence, we identify $T_{c2}$ as the spin-Peierls
transition temperature in TiOCl.  The congruence with the basic
features of the CF theory allows us to conclude that that TiOCl is,
in fact, a good realization of a spin-Peierls system.

While the CF theory can successfully describe many aspects of the
phonon data, there are some discrepancies.  Such discrepancies are
not completely unexpected, since the theory is based on a mean-field
RPA treatment of the phonons in which fluctuation corrections due to
the phonon dynamics are absent.  As noted above, the best fit $g$ is
temperature dependent and the fitted value of $J$ is a factor of 3
lower than that estimated from the magnetic susceptibility. This
likely stems from limits to the applicable range for
$\Pi_{CF}(q,\omega)$, which is most accurate in the vicinity of
$q\approx0.5$ and for temperatures near $T_{SP}$. In order to more
fully describe the data, additional anharmonic effects which broaden
the phonon lineshapes away from $q=0.5$ must be included. In
addition, we observe the zone boundary phonon remains strongly
over-damped even at temperatures as high as $T=300$~K ($\gg
T_{SP}$).  These unusual features of the phonon dynamics of TiOCl
require better theoretical understanding.

\section{Discussion and Summary}

The TiOCl compound has many ingredients which make it a particularly
ideal spin-Peierls system. First, it is composed of weakly
interacting $S=1/2$ spin chains which are well described by a
dominant nearest-neighbor interaction along the
chain.\cite{Seidel03} The weak magnetic coupling between chains is
facilitated by the staggered arrangement of adjacent Ti chains.
Indeed, a lattice dimerization is observed at low temperatures,
concomitant with spin-gap behavior observed in the susceptibility.
However, as a consequence of the staggered Ti chains, an
incommensurately modulated phase develops prior to the commensurate
dimerization upon cooling.  Our detailed x-ray scattering studies
have led to considerable new insight into this incommensurate phase.
We have developed a model which describes how the Ti chains are
modulated relative to each other and shows good agreement with the
data.  Our analysis of the higher harmonic scattering indicates a
cross-over from sinusoidal to a solition-like modulation as the
temperature is reduced below $T_{c2}=92$~K.  The
incommensurate-to-commensurate transition at $T_{c1}=66$~K is
characterized by a divergence of the domain wall spacing and
selection of a single dimerized domain.

TiOCl is also of interest since it is an inorganic material for
which single crystals samples can be grown.  Using single crystal
samples, we have performed high-resolution inelastic x-ray
scattering measurements of the lattice dynamics.  This is a
noteworthy example in which x-rays rather than neutrons can provide
the best information on the detailed lattice dynamics of a
correlated electron system.  We have discovered a soft phonon mode
which drives the spin-Peierls transition. This stands in marked
contrast to the much studied CuGeO$_3$ system, for which a soft
phonon has not been observed. In TiOCl, the lattice dynamics suggest
that the spin-Peierls temperature is associated with the transition
temperature $T_{c2}=92$~K, and not $T_{c1}=66$~K as has been
previously speculated.

Our measurements on TiOCl allow to make an unprecedently detailed
comparison with the Cross-Fisher theory.  We find that the
calculated polarizability can successfully describe the measured
phonon cross sections.  That is, the extracted values for the bare
phonon frequencies and spin-phonon coupling give a consistent
picture of both the phonon dispersion and the phonon softening at
$T_{SP}$. However, several discrepancies exists, such as the small
value of $J$ and a temperature dependent spin-phonon coupling
constant. This points to the necessity of adding additional
anharmonic effects to fully describe the lattice dynamics.
Interestingly, the soft phonon mode remains strongly overdamped at
temperatures much higher than the observed spin-Peierls transition
temperature.

The presence of the soft mode transition indicates that TiOCl falls
within the adiabatic regime (i.e., small $\Omega_0$).  Indeed, the
fitted bare phonon frequency of the Peierls-active mode,
$\Omega_0\simeq 27$~meV, is smaller than the magnetic energy scale
$J\simeq 57$~meV in the system.  Recent theories have expanded upon
the Cross-Fisher result and shed further light on the conditions
separating the adiabatic and anti-adiabatic
regimes.\cite{Gros98,Holicki,Orignac04,Citro} Gros and Werner have
shown that within RPA a soft phonon occurs only if $\Omega_0/T_{SP}
< 2.2$.\cite{Gros98} For TiOCl, we calculate $\Omega_0/T_c \simeq
2.9$ (using $T_{c2}=92$~K as the critical temperature). Hence, their
prediction suggests that TiOCl falls outside of the adiabatic
regime.  However, a more recent theory by Dobry {\em et
al.}\cite{Dobry07} goes beyond RPA and takes into account the
dynamics of the transverse phonons arising from inter-chain
interactions. They find that soft phonon behavior may occur for
values as high as $\Omega_0/T_c \simeq 3$, provided that
$\omega_\perp/T_{SP}$ for the transverse phonon is not very
large\cite{Dobry07}.  Our shell model calculations indicate that the
latter condition is satisfied in TiOCl.  Of course, further
experimental studies of the transverse phonon dynamics would
certainly be useful.  Also, more theoretical work based explicitly
on the structure of staggered Ti chains would be necessary for
additional detailed comparisons.  With its wealth of observed
phenomena and relatively simple crystal structure, TiOCl is an ideal
system for quantitative tests of theories for coupled spin and
lattice degrees of freedom in one-dimensional magnets.

\begin{acknowledgments}
We thank P. A. Lee, T. Senthil, P. Ghaemi, T. Yildirim, and J. Hill
for useful discussions. The work at MIT was supported by the
Department of Energy under grant number DE-FG02-04ER46134.
\end{acknowledgments}

\

$^*$ email: younglee@mit.edu

\bibliography{tiocl_xray}

\end{document}